\documentclass[11pt,twoside]{article}

\usepackage{myasp2004}
\usepackage{epsf}
\usepackage{lscape}
\usepackage{graphicx}

\begin{document}
\title{Results from the Exoplanet Search Programmes with BEST and TEST}   
\author{Jochen Eisl\"offel$^1$, Artie P. Hatzes$^1$, Heike Rauer$^2$, 
Holger Voss$^1$, Anders Erikson$^2$, Philipp Eigm\"uller$^1$, Eike Guenther$^1$}   
\affil{$^1$ Th\"uringer Landessternwarte, Sternwarte 5, D-07778 Tautenburg,
  Germany\\ 
$^2$ Deutsches Zentrum f\"ur Luft und Raumfahrt, Institut f\"ur
  Planetenforschung, Rutherfordstra{\ss}e 2, D-12489 Berlin-Adlershof, Germany}    

\begin{abstract} 
Th\"uringer Landessternwarte Tautenburg (TLS) has started to operate a small
dedicated telescope - the Tautenburg Exoplanet Search Telescope (TEST) -
searching for transits of extrasolar planets in photometric time series
observations. In a joint effort with the Berlin Exoplanet Search Telescope
(BEST) operated by the Institut f\"ur Planetenforschung of the "Deutsches
Zentrum f\"ur Luft- und Raumfahrt (DLR)" at the Observatoire de Haute-Provence
(OHP), France, two observing sites are used to optimise transit search. Here,
we give a short overview of these systems and the data analysis. We describe 
a software pipeline that we have set up to identify transit events of
extrasolar planets and variable stars in time series data from these and other 
telescopes, and report on some first results. 
\end{abstract}

\section{Introduction}
With the discovery of a planet orbiting the star 51\,Peg \cite{mq95} the field
of extrasolar planet research has become one of the ``hot topics'' in
astrophysics. More than 170 such extrasolar planets are known today. They
show us that other planetary systems can greatly differ from our own solar
system. The vast majority of these planets has been detected with precise
radial velocity measurements. These measurements unveil the periodic Doppler
shift in the spectral lines of the planet host star, caused by the orbital
motion. 

More recently, also transits of extrasolar planets across the disk of their
host star have been observed. These observations are of particular interest:
they allow us to measure the radii and orbital inclinations of the planets, 
and thus
-- together with radial velocity measurements -- their masses and densities. 
Therefore, numerous searches for transiting extrasolar planets have been
started in recent years, with optics from standard telephoto lenses up to 4-m
class telescopes with wide-field mosaic CCD arrays \cite{h03}. While these 
experiments are mainly geared towards the detection of Jupiter-like gas giants
in short period orbits, future space missions -- like CoRoT operated by CNES 
and Kepler by NASA -- even aim at the detection of small terrestrial planets. 

One should note, however, that high precision radial velocity measurements
require a large number of photons. Therefore, such follow-up is  only 
possible for planets
of brighter host stars. This fact becomes very obvious when looking at the
currently known transiting planets: most of them were found by the OGLE-III 
experiment \cite{upzsksswp}, a 1.3-m telescope operated at Las Campanas 
Observatory by Warsaw University Observatory, Carnegie Institution of
Washington and Princeton University. Their stars have typical brightnesses 
of V = 14 -- 17. Even with long
integrations at the ESO VLT and Keck telescopes the achievable precision for
radial velocity measurements is such, that they limit the accuracy with which
physical parameters of these transiting planets and their orbits can be
determined \cite{bummpidlqssz}. 

Photometric transit searches with smaller wide-angle optics may therefore have
the great advantage of finding transiting planets around brighter stars. For
these, then detailed studies of their atmospheres become possible. The
detection of Na and H, C, O in the atmosphere and exosphere of HD\,209458
\cite{cbng01, vldbfhm03, vdlhbefmmp04} 
with the Hubble Space Telescope and the measurements of the albedo of
HD\,209458 and TrES-1 with the Spitzer Space Telescope 
\cite{camtabglmos05, dsrh05} are examples of what information can be obtained
for planets with brighter host stars. 
Therefore, Th\"uringer Landessternwarte Tautenburg (TLS) and the Institut
f\"ur Planetenforschung of the "Deutsches Zentrum f\"ur Luft- und Raumfahrt
(DLR)" have decided to operate such small (20 -- 30\,cm) telescopes to
search for planetary transits of bright host stars. In the course of these
measurements also a wealth of variable stars are found. 

In Sect.\,2 we describe the BEST and TEST telescopes. In Sect.\,3 we discuss
our the analysis pipeline that we have set up at TLS to find transit events
and variable stars in photometric time series data. Finally, we show some
early results in Sect.\,4.

\begin{figure}[th]
\begin{center}
\includegraphics[width=0.60\textwidth,angle=-90]{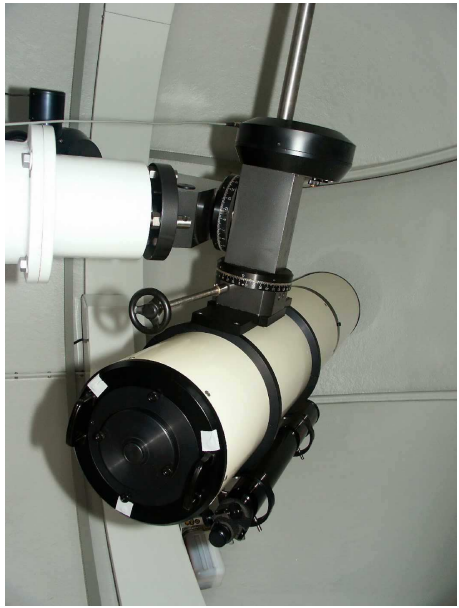}
\end{center}
\caption{The Tautenburg Exoplanet Search Telescope TEST in its dome.}
\end{figure}

\section{The BEST and the TEST}

From 2001 to 2003 DLR operated the BEST telescope at the Th\"uringer
Landesternwarte in close collaboration with TLS. BEST is a Flatfield Camera of
Schmidt-type with 20\,cm aperture and 54\,cm focal length
\cite{reeghmv04}. It was equipped with an APOGEE CCD camera with 2k$\times$2k
pixels giving a field of view of 3.1$\times$3.1 deg$^2$. This telescope has
been moved to the Observatoire de Haute Provence (OHP) in 2004, where it is
operated from Berlin and used mainly as support facility for the CoRoT space
mission. In Tautenburg, a replacement was installed in a newly constructed
dome in the course of 2005. This new 
telescope -- the Tautenburg Exoplanet Search Telescope, or TEST -- is a larger 
version of Flatfield camera with 30\,cm aperture and 94\,cm focal length (see 
Fig.\,1). It is equipped with an APOGEE CCD camera with 4k$\times$4k
pixels with a scale of 2.0\,arcsec/pixel, giving a field of view of
2.2$\times$2.2 deg$^2$. Currently, the BEST is operated in white light, while
the TEST observes through an R-band filter. Both telescopes may be equipped 
with filter wheels for complete filter sets in the future. These, however, 
have to be custom-build, since commercial filter wheel have too small clear 
apertures, and would vignette a large part of the field.

\begin{figure}[th]
\begin{center}
\includegraphics[width=0.75\textwidth,angle=-90]{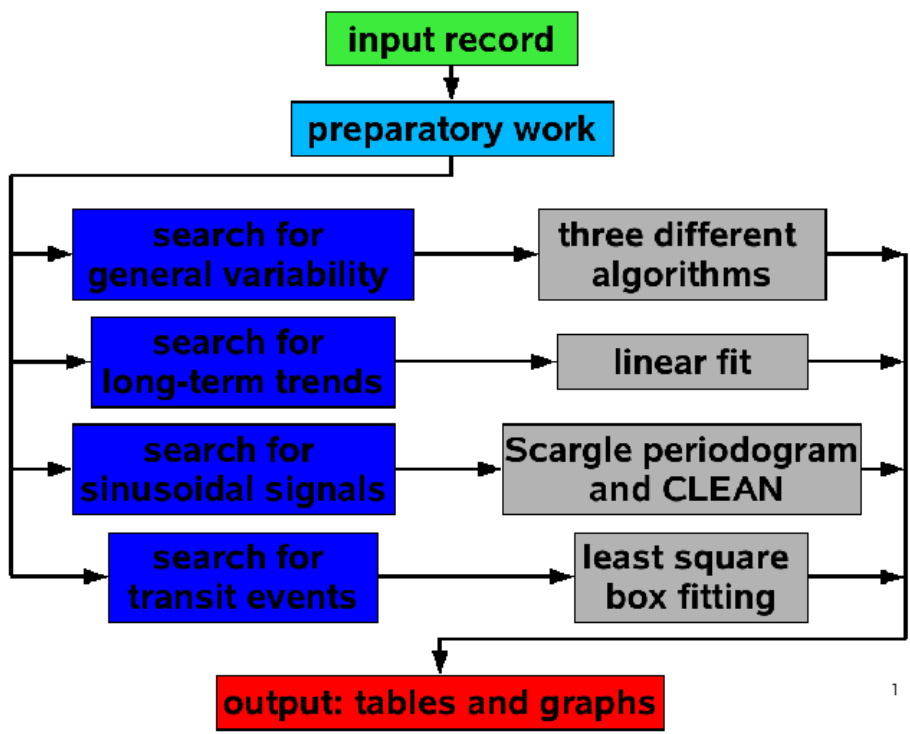}
\end{center}
\caption{Schematic of the detection pipeline that we have set up at TLS to
  identify variable objects in photometric time series data sets.}
\end{figure}

\section{Searching for transiting planets and variable stars}

With both telescopes selected fields in the vicinity of the Galactic plane
with high stellar density are being monitored throughout a whole observing
season. Furthermore, we obtained extended photometric time series of fields in
young open clusters in the course of a programme to study rotation periods of
very low-mass stars and brown dwarfs with various other telescopes
\cite{se04a, se04b, se05}, which we now are searching for planetary transits
as well. Currently, target fields of the CoRoT satellite mission are being
observed with BEST at the OHP \cite{krev06}.

All the images from these various time series observations are reduced
following standard recipes, including bias subtraction, flat-fielding, and
fringe correction where necessary. Then, photometry of all sources on these
images is carried out. In the beginning, aperture photometry was used for the
BEST data \cite{reeghmv04}. This has been upgraded to differential
image analysis in the meantime \cite{krev06}. On the images of the cluster 
fields from the rotation project, on the other hand, PSF (point spread 
function) photometry was done using daophot \cite{s87}. 

In all cases the data were then detrended on a nightly basis. 
We describe in detail how this is being done for the cluster fields from the
rotation project in Scholz and Eisl\"offel (2004a, 2004b, 2005) using
non-variable stars in the same fields (see Rauer et al. 2004 for the BEST 
data). Atmospheric effects like extinction from changing
airmass and cirrus are effectively corrected with this approach. 

We have then set up a data analysis pipeline to search such detrended
photometric time series data sets for transits of extrasolar planets and
variable stars \cite{e06}. While these routines have been developed using data
from the rotation project and the BEST fields, this pipeline should be capable
of handling data from other time series observations equally well. 

Since we are searching for different kinds of variable signals, a variety of
different algorithms have to be used (see Fig.\,2). Initially, a general
search for photometric variability is carried out. This step employs three
different variability indicators: a) the modified Stetson-Welch variability
index \cite{ws93, zdxz03}, b) the standard deviation within each lightcurve 
as a function of the object brightness, c) a comparison of the mean 
brightness and standard deviation of fractions (halfs, thirds, quarters ...) 
of the light curve. 

For certain types of variability more specialised algorithms may deliver
better detection efficiencies, and at the same time permit the determination
of parameters of these signals that allow us already a crude classification of
the variable objects. Some objects may exhibit variability on such long time
scales, that within our observing window only a linear trend can be observed
in their light curves. Others may show periodic signals. Pulsating, rotating,
and eclipsing systems with periods shorter than the observing window belong to
this group. Although many of their light curves are not strictly sinusoidal,
algorithms tailored to the detection of such signals manage to identify those
objects very well. We implemented the Scargle-Lomb periodogram search
\cite{s82}, followed by the CLEAN algorithm \cite{rld87} for this purpose. 
Then, we determine the significance of the found periods. 

Signals for which the duration of variability is short compared to the period
length, such as Algol systems and transits of extrasolar planets, are usually 
not found by these algorithms. These events are instead very favourably 
detected by the box least-square fitting (BLS) algorithm developed by Kovacs et
al. \cite{kzm02}. The significance of the detected events is then determined
using the signal detection efficieny (SDE). In addition, we can subtract 
long-term trends
or sinusoidal signals that were detected in the earlier steps of the analysis
from the light curves in order to increase the sensitivity of the BLS search.

\begin{figure}[th]
\begin{center}
\includegraphics[width=0.45\textwidth,angle=-90]{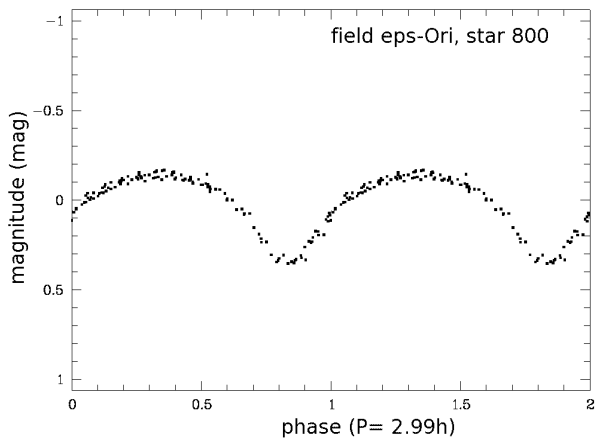}
\end{center}
\caption{An eclipsing binary found with the TLS detection pipeline for
  variable  objects in a cluster field near $\epsilon$ Orionis.}
\end{figure}

\section{Early results}

We have found many variable objects in the fields from our rotation project
and in BEST fields that we searched with our analysis pipeline. 
In Fig.\,3 we show the phased light curve of one of the contact binary systems
that we found in one of our time series fields from the rotation project 
of the young cluster near $\epsilon$ Orionis (although this is most likely 
a background system, not related to this young cluster itself) \cite{e06}.

In the fields that were observed with the BEST while it was operated at TLS,
several candidates for transiting extrasolar planets were found. Until now,
however, follow-up spectroscopy has identified most of these candidates as
eclipsing binary systems. One of these, GSC 3566-1556 (see Fig.\,4), was found
to be comprised of a G0V-type primary and a M3V-type secondary, causing an
eclipse of a depth similar to a planetary transit \cite{reeghmv04}. 

The lightcurves obtained with BEST while operated by DLR at TLS will be made
available to the community (contact H. Rauer if interested). The data obtained
on the CoRoT target fields will be added to the CoRoT ExoData data base.

\begin{figure}[th]
\begin{center}
\includegraphics[width=0.33\textwidth,angle=-90]{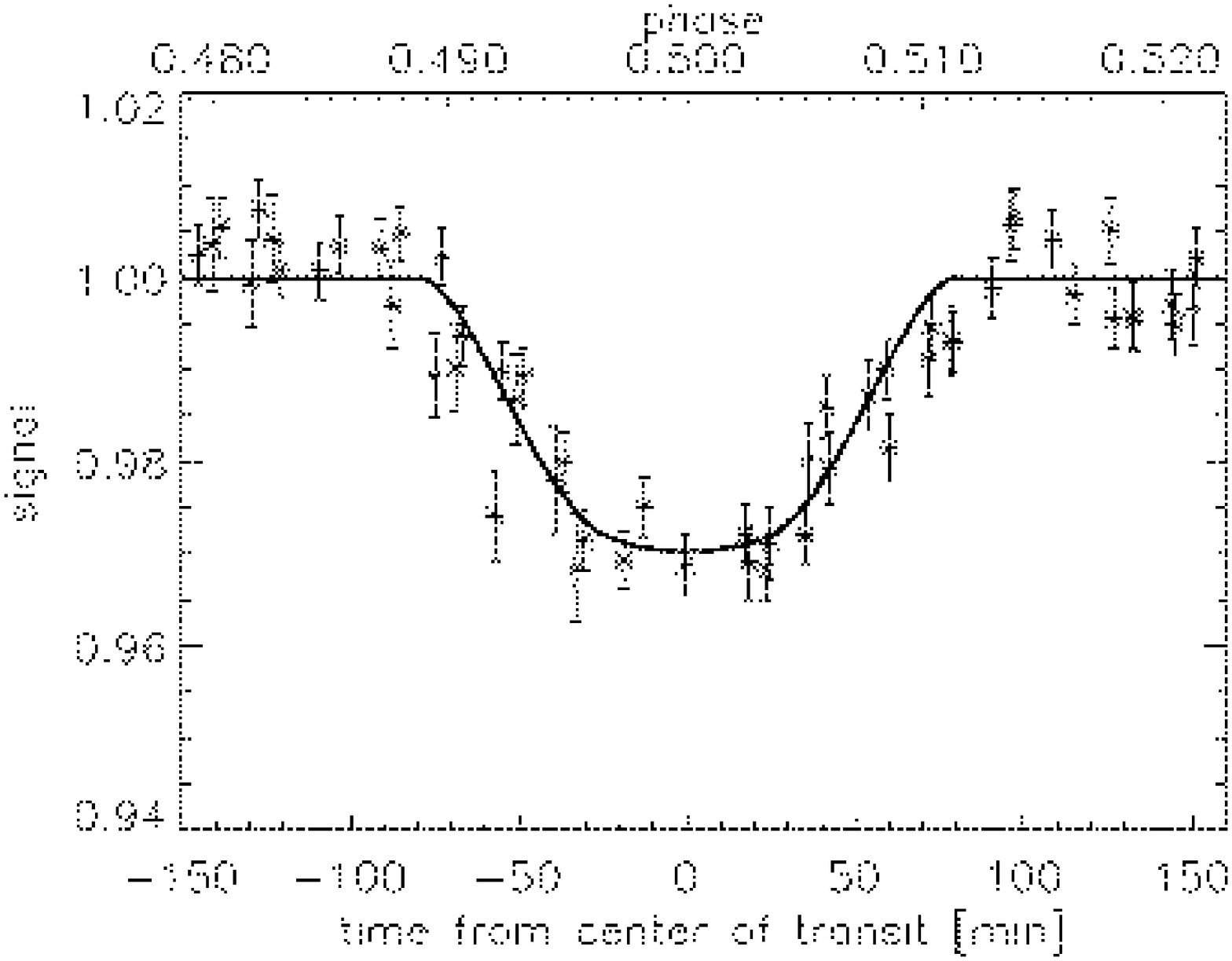}
\includegraphics[width=0.33\textwidth,angle=-90]{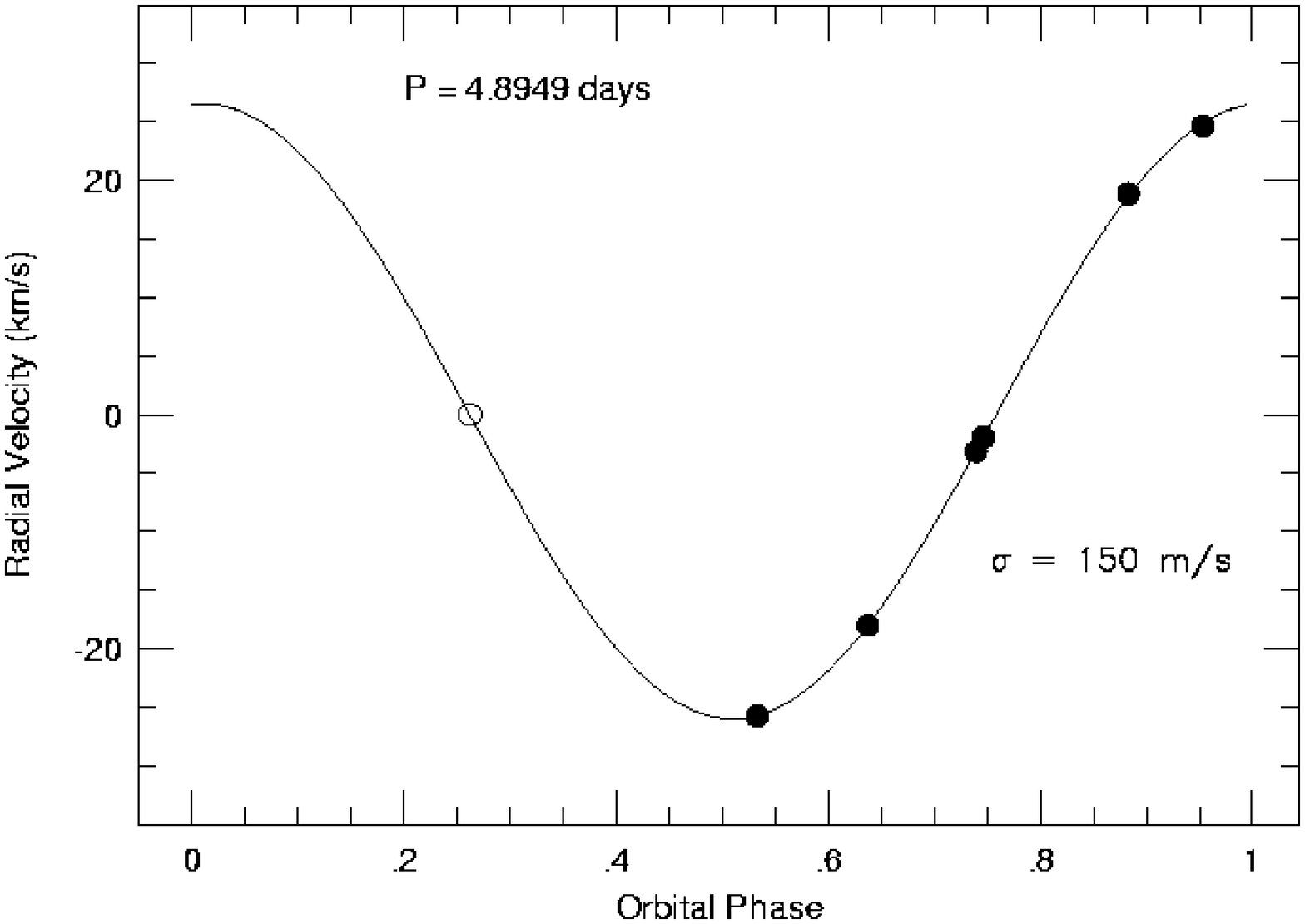}
\end{center}
\caption{Left panel: combined light curve of the BEST transit candidate GSC
  3566-1556 \cite{reeghmv04}. Right panel: New radial velocity measurements of
  this candidate obtained with the TLS 2-m telescope. The open circle shows the
  orbital phase when the transit events were observed.}
\end{figure}

\acknowledgements
This work was partially supported by German Verbundforschungsgrant 50OW0204
from the Deutsches Zentrum f\"ur Luft- und Raumfahrt e.V. (DLR).


\end{document}